# Fermi-Dirac Statistics Applied to Very Dense Plasmas at Medium or Low Temperatures with Optical Parameters Calculations

Y.Ben-Aryeh [*]

Technion-Israel Institute of Technology, Physics, Haifa 32000, Israel

[*] Corresponding author: Y. Ben-Aryeh   phr65yb@physics.technion.ac.il

Fermi Dirac free electron model is applied to very dense plasmas with medium or low temperatures. While Boltzmann statistics can lead to very high densities of ionized electrons, only at very high temperature, Fermi Dirac statistics can support the high densities of ionized electrons at medium or low temperatures due to the high degeneracies obtained in this model. Since very dense plasmas may be obtained at low temperatures the corresponding black body radiation with the plasma luminosity will be quite small. On the other hand gravitational effects might be quite large due to the high densities. The plasma optical parameters for dense plasmas are calculated. The present study might have implications to dense stars plasma.

## 1. INTRODUCTION

Optical properties of plasmas are treated in various books (see e.g. [1-7]). A plasma matter is defined as a mixture of electrons and positive ions which is neutral, but deviations from neutrality can develop due to perturbations including interactions with electro-magnetic fields among others. Usually [1-7] one uses Boltzmann statistics for describing the statistics of ionized electrons but we would like to treat here the plasmas under the conditions in which quantum effects become important and Boltzmann statistics is exchanged to Fermi Dirac statistics.

By taking into account that the ratio $M_i/m_{el}$ is very large, where $M_i$ and $m_{el}$ are the ion and the electron mass, respectively, the plasma properties are described mainly by the electrons which are very mobile relative to the ions. The electron De Broglie wave length is given by

$$\lambda_{DB} = \frac{h}{m_{el} v_{el}} \quad . \tag{1}$$

Here $h$ is the Planck constant, $m_{el}$ and $v_{el}$ are the mass and velocity of the electron, respectively, and the quantum effects become important when the De Broglie wave length is large relative to the average distance between the ionized electrons. Putting some numbers according to Boltzmann statistics as $\frac{1}{2}m_{el}v_{el}^2 = \frac{3}{2}k_B T$ we get

$$v_{el} = \sqrt{3k_B T / m_{el}} \simeq 0.67432\sqrt{T} \cdot 10^4 \ (m \cdot \sec^{-1}) \qquad . \tag{2}$$

The De Broglie wave length is then given by

$$\lambda_{DB} \simeq \frac{1.0788}{\sqrt{T}} \cdot 10^{-7} \ (m) \qquad . \tag{3}$$

If $\lambda_{DB} \gg n_{el}^{-1/3}$, where $n_{el} \ (m^{-3})$ is the number of electrons per unit volume, and $n_{el}^{-1/3} \ (m)$ is the average distance between the electrons, then Boltzmann statistics breaks down and it should be exchanged to Fermi-Dirac statistics. Assuming, in an example $T = 4000 \ ^0K$, then the De Broglie wavelength according to Boltzmann statistics is given by $1.7056 \cdot 10^{-9} \ (m) = 17.056 \ A^0$. Therefore, we should exchange, for this example, the Boltzmann statistics to Fermi-Dirac Statistics when $n_{el} > 10^{27}$. The transition from Boltzmann statistics to Fermi Dirac statistics becomes stronger when the temperature is decreased and/or the electron density is increased.

Fermi-Dirac statistics was found to be very important for the physics of white dwarf stars (see e.g. [8]). White dwarf stars are compact objects with extremely huge densities which are the common end product in the evolution of many stars. It was shown that the electrons in these stars obey the Fermi-Dirac statistics with high degeneracies (related to Pauli-exclusion principle) which prevent their collapse after their fuel was finished. In the present work we treat plasmas with densities much smaller than those in the white dwarfs but still under the condition that Fermi-Dirac statistics should be used. The degeneracies obtained by the present Fermi Dirac free electron model are found to have important effects on the plasma optical properties.

The emissivity of gases at moderate high temperatures, of about $1000 - 2000, ^oK$, was treated by a statistical model which takes into account the spectra of these hot gases (see e. g. [9-10]). In these works Boltzmann statistics was used as the densities of these gases were relatively low. In the present work we treat ionized gases of light atoms at medium or low temperature, where very high ionized electron densities are obtained related to the Fermi-Dirac free electron statistics. We discuss the special optical properties obtained of such plasmas relating their properties to Fermi-Dirac statistics. We give detailed calculations in an example in which, the density of ionized electrons and the plasma temperature are given, respectively. We would like to emphasize, however, that this model can be applied for lower temperatures where the densities of these gases are enough high so that Fermi Dirac statistics with the high degeneracies should be used.

For the very dense plasmas and relatively low temperatures we find that the collision frequency might be larger than the optical or infra-red frequencies. Most analyses on plasmas assume that the frequency of collisions is smaller than the optical frequency (see e.g. [11-13]) but in our analysis we treat the other condition $q > \omega$ where $q$ and $\omega$ are the collision and optical frequency, respectively. While other works (see e.g. [11-13]) treated the effect of a magnetic field on the plasma transparency we treat the effect of high rate of collisions on the optical parameters of the plasma.

The present paper is arranged as follows: In section 2 we develop Fermi Dirac free electron model [14-17] in which the Fermi-Dirac statistics is applied to dense plasma. In section 3 the optical properties of the plasma with very high frequency of collisions are calculated in presence of monochromatic electromagnetic field. We develop explicit calculations in the present example in which, the density of ionized electrons and the plasma temperature are given by $n_{el} = 10^{29} \left( m^{-3} \right)$ and $T = 4000\ ^0 K$, respectively. In section 4 we summarize our results and conclusions.

## 2. FERMI-DIRAC FREE ELECTRON MODEL WITH THE NON-RELATIVISTIC APPROXIMATIONS DEVELOPED FOR PLASMAS WITH VERY HIGH DENSITIES

The total number of electrons, $N$ in the non-relativistic approximation is given by the Fermi-Dirac equation [14-17]:

$$N = 2 \frac{4\pi V}{h^3} \int_0^\infty f(x,p) p^2 dp \qquad . \qquad (4)$$

Here $V$ is the volume of our system and $N/V = n$ is the number of electrons per unit volume. $f(\vec{x}, \vec{p})$ is the Fermi-Dirac probability density function, i.e. the probability of finding a particle at position between $\vec{x}$, and $\vec{x}+d\vec{x}$, and its 3-momentum between $\vec{p}$, and $\vec{p}+d\vec{p}$, $E_{el} = p^2/2mkT$ is the electron non-relativistic energy in units of $kT$ where $k$ is the Boltzmann constant, $T$ is the temperature and $\mu$ is the Fermi-Dirac constant. $h^3$, is the volume of a quantum cell in phase space where $h$ is the Planck constant and the factor 2 follows from the two spin states of the electron. The Fermi-Dirac constant $\mu$ is to be chosen in the particular problem in such a way that the total number of electrons comes out correctly-that is equal $N$.

For complex degenerate fermions with $\mu/kT = \eta \gg 1$ the parameter $\mu$ is referred as the Fermi energy $E_f = \dfrac{p_F^2}{2m}$ and then the function $f(\vec{x}, \vec{p})$ is given by

$$f(\vec{x},\vec{p}) = f(E_{el}) = \begin{cases} 1 & \text{for } E_{el} = \dfrac{p^2}{2m} \leq E_f \equiv \mu = \dfrac{p_F^2}{2m} \\ 0 & \text{for } E_{el} = \dfrac{p^2}{2m} > E_F \equiv \mu = \dfrac{p_F^2}{2m} \end{cases} \quad ; \quad \eta = \dfrac{\mu}{kT} \gg 1 \quad . \qquad (5)$$

By substituting the approximation (5) into eq. (4) we get [16]

$$n = N/V = \int_0^{p_F} \dfrac{8\pi p^2}{h^3} dp = \dfrac{8\pi p_F^3}{3h^3} \quad , \quad p_F = h\left(\dfrac{3n}{8\pi}\right)^{1/3} = \sqrt{2mE_F} \quad . \qquad (6)$$

(Eq. (6) is equivalent to eq. (33) in [16] chapter 7).

We showed, in the introduction, for the example in which: $n \simeq 10^{29} \, (m^{-3}), T = 4 \cdot 10^{3} \, {}^{0}K$ that Boltzmann statistics break down and we need to use the Fermi-Dirac statistics. According to eq. (6) we have, in this example, the relations:

$$10^{29} = \frac{8\pi p_F^3}{3h^3} \simeq 2.8806 \cdot 10^{100} p_F^3 \rightarrow p_F \simeq 1.514 \cdot 10^{-24}\, Kg \cdot m \cdot \sec^{-1}$$
$$\eta = \frac{p_F^2}{2mkT} \simeq 22.8 \gg 1 \tag{7}$$

We would like to compare the approximate equation (6) for the electron density $n$ with the more accurate eq. (4).

The integral over $f(x,p)$ in eq. (4) can be transformed by using the following exchange of variables:

$$\varepsilon = p^2 / 2mkT \quad ; \quad dp = \frac{mkT}{p}d\varepsilon = \varepsilon^{-1/2}\frac{(mkT)^{1/2}}{\sqrt{2}}d\varepsilon \quad ; \quad \frac{\mu}{kT} = \eta \tag{8}$$

Then $n = N/V$ of eq. (4) is obtained as

$$n = \frac{8\pi}{h^3}\int_0^\infty \frac{p^2}{\exp\left[(\frac{p^2}{2m}-\mu)/kT+1\right]}dp = \frac{4\pi}{h^3}(2mkT)^{3/2}\int_0^\infty \frac{\varepsilon^{1/2}}{1+\exp(\varepsilon-\eta)}d\varepsilon =$$
$$5.446\cdot 10^{21}(T)^{3/2}\int_0^\infty \frac{\varepsilon^{1/2}}{1+\exp(\varepsilon-\eta)}d\varepsilon \tag{9}$$

The integral $\int_0^\infty \frac{\varepsilon^{1/2}}{1+\exp(\varepsilon-\eta)}d\varepsilon$ is defined as the Fermi integral.

One should take into account that, although Fermi-Dirac free electron model is usually used in the semiconductors field [14-17], here this model is applied by the integration over all the ionized electrons. The Fermi integral of (9) is related to a table in [17] given for the function $F_{1/2}(\eta)$ as:

$$F_{1/2}(\eta) = \frac{1}{(1/2)!}\int_0^\infty \frac{\varepsilon^{1/2}}{1+\exp(\varepsilon-\eta)}d\varepsilon \quad ; \quad (1/2)! = \sqrt{\pi}/2 \simeq 0.88622 \tag{10}$$

For $\eta \gg 1$ it is approximated as [14]:

$$F_{1/2}(\mu) = \left(4\eta^{3/2}/3\pi^{1/2}\right) \tag{11}$$

By using eqs. (9-10), with the value $\eta = \dfrac{p_F{}^2}{2mkT}$ , in eq. (9) we get

$$n = \frac{4\pi}{h^3}\left(\frac{4}{3\sqrt{\pi}}\right)\cdot 0.8622(2mkT)^{3/2}\left(\frac{p_F{}^2}{2mkT}\right)^{3/2} = \frac{8\pi}{3h^3}\cdot 0.972 p_F{}^3 \quad . \tag{12}$$

We find that for $\eta \gg 1$ eq. (6) gives a quite good approximation for the density of the electrons $n$ as function of the one parameter $p_F$ as obtained by comparison of (12) with eq. (6). For cases for which $\eta$ is of order 1 or smaller the density of the electrons is function of the two parameters: $T$ and $\eta$ as described by eq. (4). In any case, the distribution of the electron momentum depends on both parameters: the temperature $T$ and the electron density $n$ . (The distribution $f(\vec{x},\vec{p})$ is described in [16], Fig. 5a in chapter 7, as function of $E_{el}/\mu = \dfrac{p^2}{2m\mu}$ , for $T = 0$ , describing the step function of eq. (5), and for $T = 50000^0 K$ ). The optical properties of dense plasma are analyzed in the next section as function of the density of electrons $n$ and the temperature $T$ .

.

## 3. THE OPTICAL PROPERTIES OF PLASMA, WITH VERY HIGH FREQUENCY OF COLLISIONS, IN THE PRESENCE OF MONOCHROMATIC ELECTROMAGNETIC FIELD

Taking into account collisions the equation of motion of electron interacting with monochromatic electromagnetic field $\vec{E}(t) = \vec{E}_0 e^{i\omega t}$ is given by

$$\ddot{\vec{r}} - q\dot{\vec{r}} = \frac{e\vec{E}_0}{m}e^{i\omega t} \quad , \tag{13}$$

where $q$ is the rate of collisions ($\sec^{-1}$). A solution of this equation is given by [7]:

$$\vec{r}(t) = \frac{e}{m(\omega^2 - iq\omega)}\vec{E}(t) \quad . \tag{14}$$

The displaced electrons contribute to the macroscopic polarization $\vec{P} = -ne\vec{r}$ :

$$\vec{P} = -\frac{ne^2}{m(\omega^2 - iq\omega)} \vec{E} \qquad . \qquad (15)$$

Then the dielectric constant is given by

$$\vec{D} = \varepsilon_0 \vec{E} + \vec{P} = \varepsilon_0 \varepsilon \vec{E} = \varepsilon_0 \left(1 - \frac{\omega_p^2}{\omega^2 - iq\omega}\right) \vec{E} \quad ; \quad \omega_p^2 = \frac{ne^2}{\varepsilon_0 m} \quad . \qquad (16)$$

Here, $\varepsilon$ is the complex dielectric constant and $\omega_p^2$ is the plasma frequency of the free electron gas. The real and imaginary of the dielectric constant: $\varepsilon = \varepsilon_1 + i\varepsilon_2$ are given by

$$\varepsilon_1(\omega) = \left(1 - \frac{\omega_p^2}{\omega^2 + q^2}\right) \quad ; \quad \varepsilon_2(\omega) = \frac{-\omega_p^2 q}{\omega(q^2 + \omega^2)} \qquad . \qquad (17)$$

In most treatments of plasma, one assumes: $q \ll \omega$, i.e. the rate of collisions is very small relative to the frequency of the electric field. We treat the plasma, under the condition $q \gg \omega$ for which we get the following approximations [1]:

$$\varepsilon_1(\omega) = \left(1 - \frac{\omega_p^2}{q^2}\right) \quad ; \quad \varepsilon_2(\omega) = \frac{-\omega_p^2}{\omega q} \qquad . \qquad (18)$$

We can use the relation

$$\sqrt{\varepsilon} = \sqrt{\varepsilon_1 + i\varepsilon_2} = n + i\kappa \qquad , \qquad (19)$$

where $n$ is the index of refraction and $\kappa$ the extinction coefficient. The index of refraction $n$ and the extinction coefficient $\kappa$ are related to $\varepsilon_1$ and $\varepsilon_2$ by [7]:

$$n^2 = \frac{\varepsilon_1}{2} + \frac{1}{2}\sqrt{\varepsilon_1^2 + \varepsilon_2^2} \quad , \quad \kappa = \frac{\varepsilon_2}{2n} \qquad . \qquad (20)$$

The extinction coefficient is linked to the absorption coefficient of Beers law (describing the exponential attenuation of the intensity of a beam propagating through the medium via the relation $I(x) = I_0 e^{-\alpha x}$ )

$$\alpha(\omega) = \frac{2\kappa(\omega)\omega}{c} \quad . \tag{21}$$

The critical parameter in the present discussion is the estimation of the parameter $q$ representing the frequency of collisions. As we are interested in estimating order of magnitudes for $q$ we simplify the calculation by assuming plasma with one ionic component, with atomic number $Z_i$. By taking into account that when the electron is colliding with ion it is deflected by long range coulomb field, we estimate the value of q by electron-ion collisions. The rate of such strong ion-electron collisions was estimated by following Rutherford formulae [1,3,18] and it is given for strong collisions by:

$$q \simeq \frac{nZ_i^4 e^4}{8\pi\varepsilon_0^2 m^2 v^3} = \frac{nZ_i^4 e^4 m}{8\pi\varepsilon_0^2 p^3} \quad . \tag{22}$$

Here $v$ is the electron velocity and we used here the relation $p = mv$. For the important case of hydrogen stars plasma Eq. (15) can be used, with $Z_i = 1$. We calculate the averaged value of $q$ by integrating it over the Fermi-Dirac distribution which takes into account quantum effects. We get for $q$ averaged value $q_{Fermi}$:

$$q_{Fermi} = C \frac{2}{h^3} \int_0^\infty \frac{1}{p^3} f(x,p) d^3 p \quad ; \quad C = \frac{Z_i^2 e^4 m}{8\pi\varepsilon_0^2} \quad . \tag{23}$$

One should take into account that the expression $\frac{2}{h^3} \int_0^\infty f(x,p) d^3 p$ represents distribution of the electrons number as function of the electrons momentum. This distribution should be multiplied by $\frac{q}{n} = \frac{C}{p^3}$ where the Fermi-Dirac averaging is done only on $1/p^3$. Using the approximation of eq. (5) we get:

$$q_{Fermi} = C \frac{2}{h^3} \int_{p_0}^{p_F} \frac{4\pi p^2}{p^3} dp = \frac{8\pi C}{h^3} \ln\left(\frac{p_F}{p_0}\right) \quad . \tag{24}$$

One should notice that $q$ of eq. (15) diverges for $p \to 0$, so we introduced a lower bound $p_0$ for the integral in (24). There are different estimates for the bound for lower values of $p$ but in any case the term $\ln\left(\dfrac{p_F}{p_0}\right) = \ln(\Lambda)$ is a small number $[1,3,18]$ (of order: 10). Then we get approximately

$$q_{Fermi} \approx \frac{Z_i^2 e^4 m}{h^3 \varepsilon_0^2} \ln(\Lambda) = 2.263 \cdot 10^{16} Z_i^2 \ln(\Lambda) \qquad . \qquad (25)$$

We find the condition $q \gg \omega$ is satisfied for optical and infra-red frequencies, so that for such frequencies the approximations (18) are satisfied.

We study now the effects of high rate of collisions in our example for which $n_{el} \simeq 10^{29} \, (m^{-3})$ $T = 4 \cdot 10^3, \, ^{0}K$. It has been shown in the introduction that for such conditions the quantum effects become important so that Boltzmann statistics should be changed to Fermi-Dirac statistics. For $n_{el} \simeq 10^{29} \, (m^{-3})$ we then get:

$$\omega_p^2 = \frac{ne^2}{\varepsilon_0 m} = \frac{10^{29} e^2}{\varepsilon_0 m} = 3.183 \cdot 10^{32} \qquad , \qquad (26)$$

Inserting the value of $q_{Fermi}$ given in eq. (25) and the value of $\omega_p^2$ given eq. (26), in eqs. (18), we get:

$$\varepsilon_1(\omega) = \left(1 - \frac{3.183 \cdot 10^{32}}{5.121 \cdot 10^{32} Z_i^4 \ln^2(\Lambda)}\right) \simeq \left(1 - \frac{0.622}{Z_i^4 \ln^2(\Lambda)}\right) \quad ;$$
$$\varepsilon_2(\omega) = \frac{3.183 \cdot 10^{32}}{\omega \cdot 2.263 \cdot 10^{16} Z_i^2 \ln(\Lambda)} \simeq \frac{1.407 \cdot 10^{16}}{\omega Z_i^2 \ln(\Lambda)} \qquad (27)$$

We find in this example that $\varepsilon_1(\omega)$ tends to have the value 1, but if the density is extremely high ($n \gg 10^{29}$), $\varepsilon_1(\omega)$ will decrease significantly, i.e. will be much smaller than 1. (Relativistic effects might then become important which is not within the present approximations). We find that $\varepsilon_2(\omega)$ might be large for optical and infra-red frequencies. In this example for optical and infra-red frequencies the absorption coefficient $\alpha(\omega)$ is quite large.

While we made the special analysis in an example for which $n_{el} \simeq 10^{29} \, (m^{-3})$, $T = 4 \cdot 10^{3} \, {}^{0}K$ ; $\eta \simeq 22.8$ we can get high densities e.g. $10^{29} \, m^{-3}$ at much lower temperatures, i. e. for which $T \ll 4 \cdot 10^{0} K$ ; $\eta \gg 22.8$. The idea is that the black body radiation which is proportional to $T^4$ will be very small for cold dense plasmas. This property might have implications for stars which have low luminosity. Since the absorption coefficient might be quite large the emissivity, which is the ratio between the plasma radiation and that of black body radiation, might be nearly 1. But the temperature might be very low for plasmas with high densities.

## 4. SUMMARY AD CONCLUSIONS

We followed in the present work the idea that for very dense plasmas for which the average distance between the electrons is small relative to their De Broglie wavelength one should use the Fermi-Dirac statistics instead of the usual Boltzmann statistics. We find in section 2 that the Fermi- Dirac distribution function $f(E_{el})$ given by Eq. (4) includes the parameter $\frac{\mu}{k_B T} \equiv \eta$ which leads to broadening of the distribution function [16]. We demonstrated in section 2 the calculation of the parameter $\eta$ in a special example for which $n_{el} \simeq 10^{29} \, (m^{-3})$, $T_e = 4 \cdot 10^{3}, \, {}^{0}K$. The parameter $\eta$ obtained in Eq. (7) shows large broadening of the distribution function. This broadening can explain the high densities, e. g. $n_{el} \simeq 10^{29} \, (m^{-3})$ obtained in relatively medium temperature. We have shown that Boltzmann statistics cannot lead to such high densities in medium temperature. We expect the effects obtained by the Fermi-Dirac free electron model to be much stronger at lower temperatures.

The basic equations for the optical properties of plasmas were reviewed. These properties are developed usually by the assumption that the optical frequency is large relative to the collision frequency. We developed in the present article the optical properties of the plasmas by assuming the opposite other extreme case for which the collision frequency is larger than the optical frequencies. The optical properties of plasmas were related to the collision frequency given by eq. (22). This collision frequency was integrated over the fermi-Dirac distribution and given in eq. (24), where the constant C is given in eq. (23). This integration was related to the approximation (4) and lower bound for the integration was related to the function $\ln(\Lambda)$.

The optical properties of a plasma for which $n_{el} \simeq 10^{29} \, (m^{-3})$ $T = 4 \cdot 10^3 \, ^0K$ were calculated, where the equations for the real and imaginary components of the dielectric constants were obtained in eq. (27). We find in this example that the real component of the dielectric constant tends to have the value 1 but the absorption coefficient given approximately by eq. (21) has quite large values.

In the present work we described very dense plasmas at medium and low temperatures according to Fermi-Dirac statistics. We found that that very high plasma densities can be obtained at low temperatures where the black body radiation which is proportional to $T^4$ is very small .The present analysis might have important implications to the study of dense stars plasmas as on the one hand we have shown that such plasmas have very weak luminescence but on the other hand such plasmas can lead to strong gravitational forces. We find that quantum effects by which very high dense plasmas have very low luminescence are related to Fermi-Dirac statistics.

The critical parameter of collision frequency $q$ was calculated in the present analysis by taking into account electron-ion collisions. It is well known that under the condition $\omega \gg q$ the effect of electron-electron collisions can be neglected. Under the condition $\omega \ll q$ the situation is less clear [1]. Since the electron-electron collision frequency depends on the electron density squared it might be that for very high densities the effect of electron-electron collisions on the optical properties of the plasma cannot be neglected. Such effects might increase further the total collisions frequency.